# A Query Language for Multi-version Data Web Archives

Marios Meimaris[1,2], George Papastefanatos[2], Stratis Viglas[3], Yannis Stavrakas[2], Christos Pateritsas[2] and Ioannis Anagnostopoulos[1]

[1]Department of Computer Science and Biomedical Informatics, University of Thessaly, Greece
janag@ucg.gr
[2]Institute for the Management of Information Systems, Research Center "Athena", Greece
{m.meimaris, gpapas, yannis, pater}@imis.athena-innovation.gr
[3]School of Informatics, University of Edinburgh, UK
sviglas@inf.ed.ac.uk

**Abstract.** The Data Web refers to the vast and rapidly increasing quantity of scientific, corporate, government and crowd-sourced data published in the form of Linked Open Data, which encourages the uniform representation of heterogeneous data items on the web and the creation of links between them. The growing availability of open linked datasets has brought forth significant new challenges regarding their proper preservation and the management of evolving information within them. In this paper, we focus on the evolution and preservation challenges related to publishing and preserving evolving linked data across time. We discuss the main problems regarding their proper modelling and querying and provide a conceptual model and a query language for modelling and retrieving evolving data along with changes affecting them. We present in details the syntax of the query language and demonstrate its functionality over a real-world use case of evolving linked dataset from the biological domain.

## 1 Introduction

The Data Web consists of an increasing quantity of scientific, corporate, government and crowd-sourced data being published and interlinked across disparate sites on the web, usually in the form of Linked Open Data (LOD). The standard way of modeling LOD is the Resource Description Framework[1] (RDF), which is a W3C recommendation. RDF supports the modelling of facts about entities in a simple triple format consisting of a subject, a predicate and an object. Entities are identified by their Uniform Resource Identifiers (URIs), which are also referred to as Internationalized Resource Identifiers (IRIs). Collections of triples form directed labelled graphs of nodes connected to other nodes or literals in semantically meaningful ways. Furthermore, the standard recommendation for querying RDF datasets is SPARQL[2], which is essentially a graph query language. Because RDF is generic enough to enable users to define custom, loose relationships between data, it is not trivial to represent more complex meta-correlations, enable annotations in data at the triple level, assign context, model changes and so on. Data-aware practices, such as data interlinking between heterogeneous sources and data visualization, have a huge potential to create insights and additional value across several sectors, however little attention has been given to the long-term accessibility and usability of open datasets in the Data Web. Linked open datasets are subject to frequent changes in the encoded facts, in their structure, or the data collection process itself. Most changes are performed and managed under no centralized administration, eventually inducing several inconsistencies across interlinked datasets. LOD should be preserved by keeping them constantly accessible and integrated into a well-designed framework for evolving datasets that offers functionality for versioning, provenance tracking, change detection and quality control while at the same time provides efficient ways for querying the data both statically and across time.

[1] http://www.w3.org/RDF/

Most of the challenges related to the management of LOD evolution stem from the decentralized nature of the publication, curation and evolution of interdependent datasets, with rich semantics and structural constraints, across multiple disparate sites. Traditional database versioning imposes that data and evolution management take place within well-defined environments where change operations and data dependencies can be monitored and handled. On the other hand, web and digital preservation techniques assume that preservation subjects, such as web pages, are plain digital assets that are collected (usually via a crawling mechanism), time stamped and archived for future reference. In contrast to these two approaches, the Data Web poses new requirements for the management of evolution [18,19]. Observe **Figure 1** where an example from the biological domain is presented. EFO is an ontology that combines parts of several life science ontologies, including anatomy, disease and chemical compounds [14]. Its purpose is to enable annotation, analysis and visualization of data related to experiments of the European Bioinformatics Institute[3]. In the figure, a URI that represents a Cell Line class changes between two consecutive versions and becomes obsolete. EFO entities are being published in LOD format, enabling other sites to reference and interlink with them. EFO is regularly updated and new versions are published on the web, usually overwriting previous ones. In this context, several interesting problems and challenges arise related to long-term preservation and accessibility of evolving LOD datasets:

***Modelling evolving datasets***. LOD datasets are evolving entities for which additional constraints may hold related to the way data is published, and evolve as dictated by domain-specific, complex changes. *This calls for appropriate modelling methods for preserving across time a multitude of dimensions related to the internal structure of a dataset, its content and semantics as well as the context of its publication*. Preservation should exhibit format-independence, data traceability and reproducibility and a common representation for data that originate from different models. Reference schemes (URIs) must be properly assigned such that unique identification and resolution is achieved across different sites, and most importantly across time. Provenance metadata can capture dataset lineage from the dataset to the record level. Distributed replication of LOD enhanced with temporal and provenance annotations can enable long-term availability and trust.

[2] http://www.w3.org/TR/rdf-sparql-query/

[3] http://www.ebi.ac.uk/

***Change management***. Changes can occur at different granularity levels. At the dataset level, datasets are added, republished, or even removed, without versioning or preservation control; at the schema level, the structure may change calling for repair and validation on new versions; finally, at the instance level data resources and facts are added, deleted or updated. Discovering changes [20] and representing them as first class citizens with structural, semantic, temporal and provenance information *is vital in various tasks such as the synchronization of autonomously developed LOD versions, or visualizing the evolution history of a particular dataset*. A unified framework that deals with evolution must be able to allow change management as a dimension of the dataset's evolution.

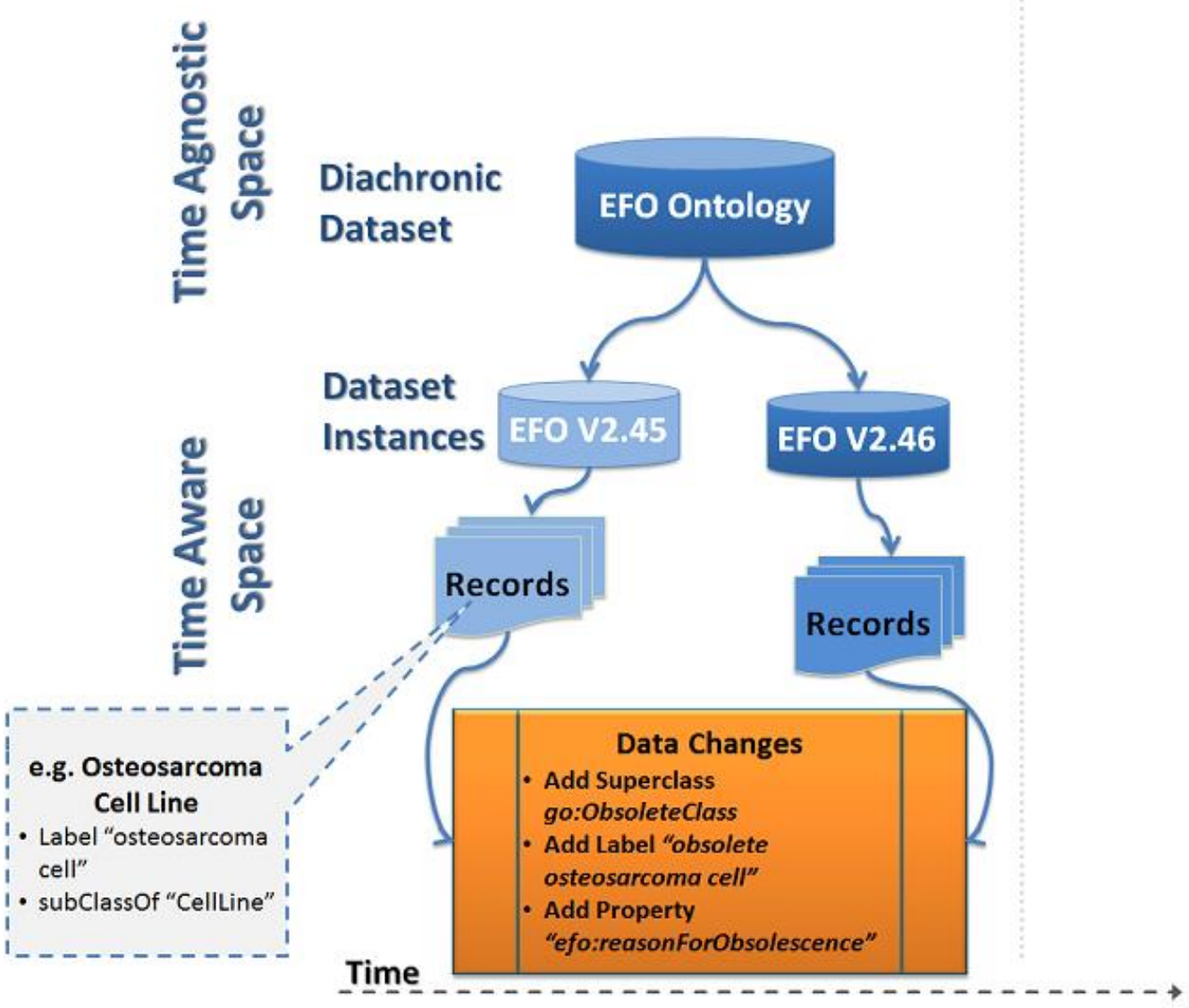

Figure 1. Evolution of a Cell Line between versions 2.45 and 2.46 of the Experimental Factor Ontology.

***Longitudinal accessibility and querying***. LOD preservation mechanisms must enable the long-term accessibility of datasets and their meaningful exploration over time. Datasets with different time and schema constraints coexist and must be uniformly accessed, retrieved and combined. Longitudinal query capabilities must be offered such that data consumers can answer several types of queries, within a version or across sets of versions. Querying must take place (i) across time, (ii) across datasets and (iii) across different levels of granularity of evolving things.

Considering the above, the benefits managing evolving LOD datasets can be placed into two categories, namely *quality control* and *data analysis*. Data evolution provides valuable insights on the dynamics of the data, their domains and the operational aspects of the communities they are found in, while tracking the history of and maintaining proper metadata of data objects across time enables better interoperability, trust and data quality.

To address these challenges, in this paper we propose a model and a query language for evolving LOD datasets. At the basis of the archive lies a conceptual model, called DIACHRON model[4] that captures structural concepts like datasets and their schemas, semantics like web resources, their properties and links between them as well as changes occurring on these concepts in different granularity levels. In the same time, our approach models in a uniform way both time-aware (evolving) and time-agnostic (diachronic) concepts, representing their between interconnections. Based on this model, a query language is designed that specifically caters for the model's inherent characteristics and takes advantage of the abstraction levels thus making the user avoid complicated, implementation-dependent queries. The query language is designed as an extension of SPARQL, specific to the DIACHRON model, that tackles the duality of data (evolving vs. diachronic objects) in order to provide a query mechanism with the ability to correlate source data with changes, annotations at various levels and other kinds of DIACHRON related metadata across time. Finally, we implement these as an archiving framework capable of storing and making available in the long term evolving LOD datasets.

This paper provides the following contributions:

1. We formally define the DIACHRON data model, a conceptual model for the representation of datasets and their evolving aspects, such as their structural, semantic, and metadata evolution. Specifically, we provide entities for modelling data that change through time in multi-version contexts, where their schema, data and metadata exhibit changes in a multitude of levels, from tuples, to collections of datasets.
2. We propose and formally define the DIACHRON Query Language as a means to enable retrieval of data and metadata across versions and datasets. The proposed query language enables querying of evolving entities across time, along with the structural elements of the entities (e.g. the reified triples) as well as the changes affecting them.
3. We provide an implementation of an archiving system that uses the DIACHRON model and implements the DIACHRON Query Language as an extension of SPARQL, and we perform experimental evaluation in terms of usability and performance on real-world datasets from the life sciences domain.

This paper is outlined as follows. In section 2 we discuss related work, in section 3 we present the DIACHRON data model, in section 4 we present the DIACHRON query language, in section 5 we describe our implementation of an archive that uses the proposed model and query language, while in section 6 we perform experimental evaluation. Finally, section 7 concludes the paper.

[4] The DIACHRON model has been informally introduced in the context of the DIACHRON project [17] and is part of the DIACHRON preservation platform, http://www.diachron-fp7.eu

## 2 Related Work

Managing LOD evolution is a multi-faceted problem that consists of versioning, efficient archiving, change representation, change detection, model abstraction and provenance tracking, among others. Work has been done in most of these fields individually, but few approaches have regarded the issue as a singular problem of many interdependencies, less so in the case of the Data Web, where datasets evolve independently, often in non-centralized ways, while citing and using one another. Versioning for LOD in the context of complete systems or frameworks has been addressed in [1,4,6,7,8,9,16,23,31]. However, these approaches address a subset of the problems discussed as will be discussed.

Ontology or schema based approaches have been proposed in [3,5,8] with the most prominent example being the PAV ontology [33], a specialization of the W3C recommended PROV ontology [12] for modelling provenance. In our work, we consider the representation of provenance as an orthogonal problem, in the sense that any model for representing metadata annotations can be used in conjunction with our work.

As far as querying is concerned, work has been done in extending SPARQL with temporal capabilities [2,11,14,23]. Contrary to our approach, in [23] no explicit data model is proposed, instead temporal information is used to separate triples in named graphs. Incorporation of annotations and provenance on the query side has been approached in [14] where triple annotations serve as context in the proposed SPARQL extension. This approach however does not differentiate between types of annotations, and is limited to treating annotations as singular tags of triples. In [2] an ontology-based approach is followed where temporal reasoning capabilities are provided to OWL-2.0 and SPARQL is extended to cater for the temporal dimension. While [2] extends an existing RDF query language with temporal reasoning, it limits its functionality in this context and does not deal with evolution of structural concepts such as datasets, tuples, or individual triples. In contrast, our approach aims at providing querying capabilities for both the semantic and the structural elements of an evolving dataset. In [12] a triple store is implemented that incorporates spatiotemporal querying by utilizing the SPARQL extensions proposed in [23]. These approaches are specifically tuned to address temporal or spatiotemporal querying in RDF data, and do not rely on conceptual models for representing in a uniform way semantically rich evolving datasets, changes, and metadata through time.

In [31] an approach is presented that builds on the Memento [30] framework, an extension of HTTP to include a traversable and queryable temporal dimension, adapted for LOD purposes. Non-changing, time-independent URIs are employed for current state identification. Dereferencing past versions of resources is done with temporal content negotiation, an HTTP extension. We draw from this work the notion of time-independent URIs for current state identification, however, we are not inter-ested in providing functionality at the HTTP level; instead we take on a data-centric rather than a document-centric approach for deep archiving and preservation of large datasets.

In [33], the authors tackle the problem of version management for XML documents by using deltas to capture differences between sequential versions and use deltas as edit scripts to yield sequential versions. The introduced space redundancy is compensated by the query efficiency of storing complete deltas rather than compressed deltas. They go on to define change detection as the computation of non-empty deltas and they argue that past version retrieval can be achieved by storing all complete deltas as well as a number of complete intermediate versions, finding the bounding versions of the desired ones and applying their corresponding deltas. Finally, they use a query language based on XQuery in order to enable longitudinal querying and they provide tag indices for each edit operation for faster delta application. While this approach deals with longitudinal querying by extending an existing standard, similar to our approach, they do not provide support for more complex semantic changes, or placeholders for capturing the evolution of other entity types, such as metadata and provenance annotations.

In [3], the authors propose a method for archiving scientific data from XML documents. The approach targets individual elements in the DOM tree of an XML document, rather than the whole versions themselves. They use time stamping in order to differentiate between the states of a particular element in different time intervals and they store each element only once in the archive. The timestamps are pushed down to the children of an element in order to reflect the changes at the corresponding level of the tree, an approach also followed in [20]. Our approach is inspired by the hierarchical attribution of time and we adopt this model and partially adapt it to the case of RDF. Moreover, we extend this hierarchical attribution to generic metadata annotations instead of strictly temporal.

In [28] the authors study the change frequency of LOD sources and the implications on dataset dynamics. They differentiate between the document-centric and the entity-centric perspectives of change dynamics, the latter further divided into the entity-per-document and global entity notions. We partially adopt this distinction in our work, as will be described further on. Specifically, we introduce a conceptual model that differentiates between entity types that represent both the structural aspects of a dataset, and the semantic ones.

SemVersion [32] computes the semantic differences as well as the structural differences between versions of the same graph but is limited to RDFS expressiveness. DSNotify [24] is an approach to deal with dataset dynamics in distributed LD. The authors identify several levels for the requirements of change dynamics, namely, vocabularies for describing dynamics, vocabularies for representing changes, protocols for change propagation and algorithms and applications for change detection. It implements a change detection framework which incorporates these points in a unified functionality scheme, having as main motivation the problem of link maintenance. Both these approaches only support full materialization of datasets, contrary to our approach that supports a hybrid model of storing datasets and semantic deltas. Fur-thermore, contrary to our approach, they do not deal with querying over time, changes and metadata.

Our approach differentiates itself by considering versioning, annotating, change management, and dataset heterogeneity as necessary components of an evolving dataset, and are thus tackled together. Furthermore, most of the work presented in this section addresses the temporal aspect of evolution in datasets, instead we chose to consider temporality as an inherent characteristic of versioning. It is trivial to explicitly create temporal operators for DIACHRON QL by evaluating datasets over their temporal metadata and translating temporal operators to version-based operators such as `AT VERSION` or `BETWEEN VERSIONS`.

## 3 An archive model for evolving datasets

Our modelling approach supports a format-independent archiving mechanism that maintains syntactic integrity by making sure that the original datasets are reproducible and at the same time takes advantage of information-rich content in these datasets. Format-independence enables different source models (e.g. relational, multidimensional, ontological) to be transformed to a common RDF representation, uniformly annotated with temporal and provenance information.

The DIACHRON model provides the basis for defining semantically richer entities that evolve with respect to their source datasets' history. At the core of the model lies the notion of the *evolving entity,* which captures both structural and semantic constructs of a dataset and acts as a common placeholder for provenance, temporal, and other types of metadata.

Evolving entities are identifiable and citable objects. These entities all share a common ancestor, the *Diachronic Entity*, which allows the aforementioned requirements to be addressed on different levels. The different types of entities in the DIACHRON model and their interactions can be seen in **Figure 2** and **Figure 3**. Specifically, **Figure 2** shows a class diagram that describes the relationships between concepts in the DIACHRON model, while **Figure 3** provides an aggregated space where concepts are partitioned in time-aware vs time-agnostic, and data (non-curated) vs curated information space. There, example instantiations between the different concepts in the data model are presented. An example drawn from the EFO ontology can be seen in **Figure 4**. The entities of the model are described in the following.

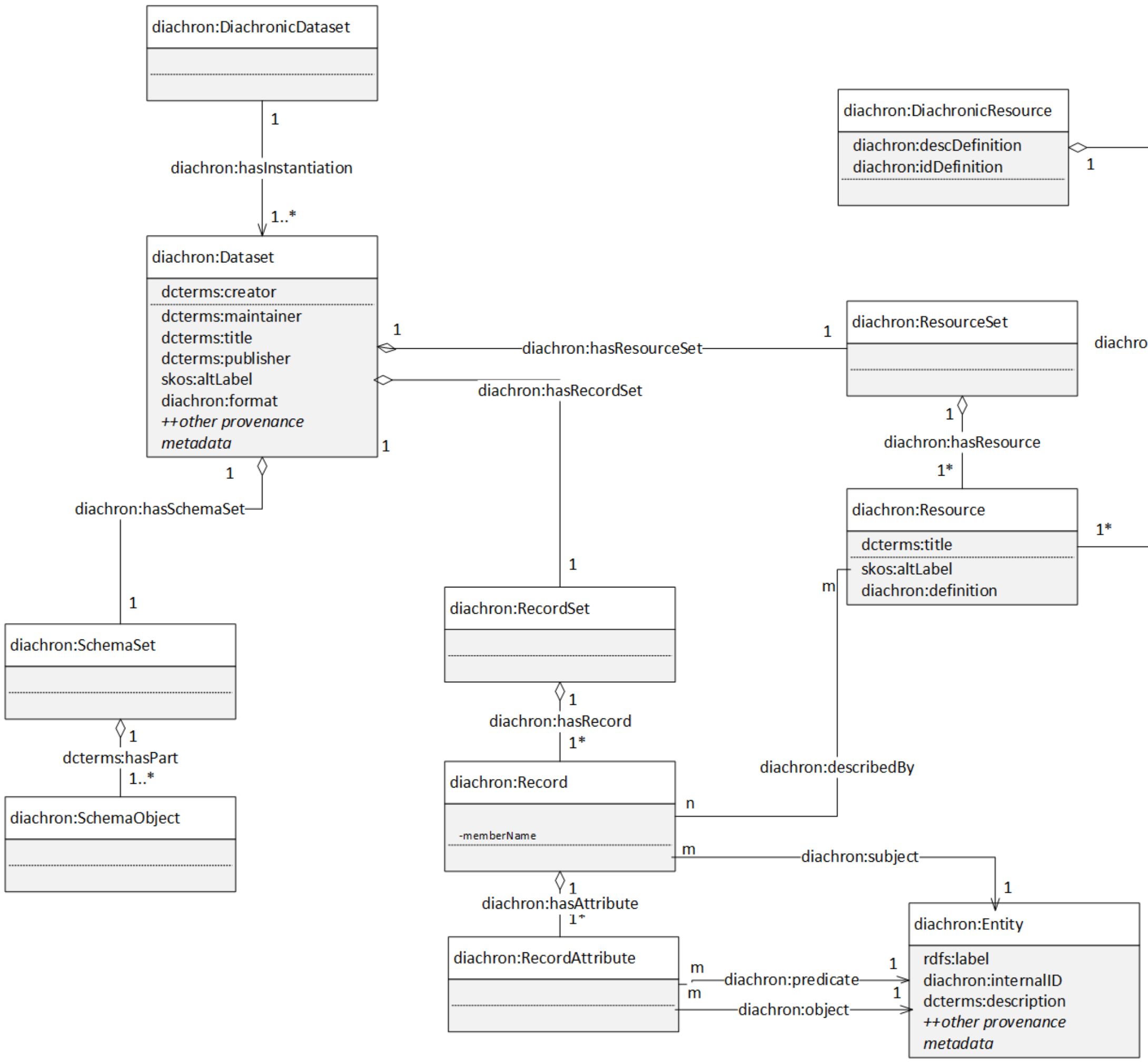

Figure 2. Class diagram for the DIACHRON model.

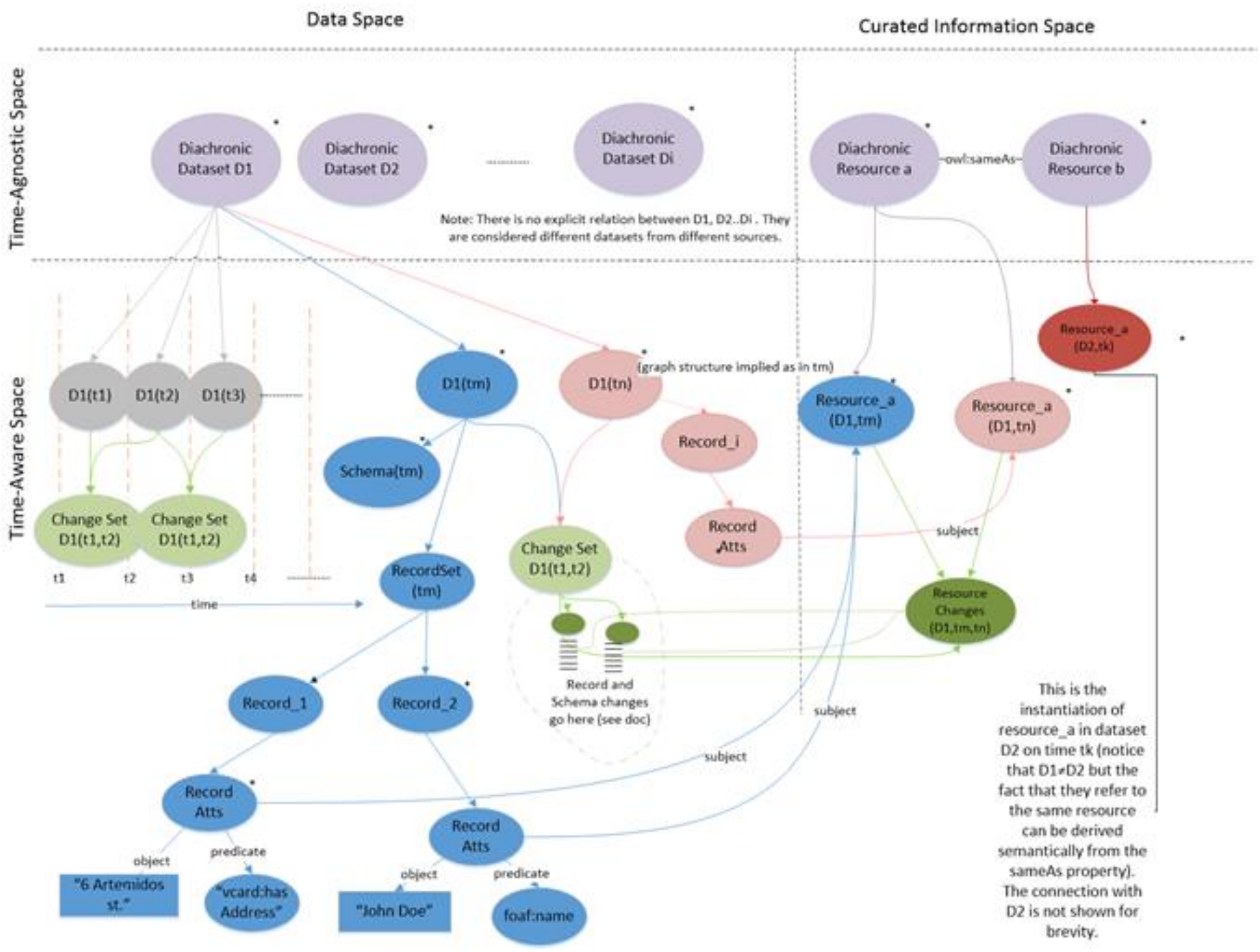

Figure 3. The DIACHRON model space.

*Diachronic datasets and dataset instantiations*. Diachronic datasets are conceptual entities that represent a particular dataset from a time-agnostic point of view, which in turn is linked to its temporal instantiations or versions. Furthermore, diachronic dataset metadata comprise information that is not subject to change, such as diachronic dataset identifiers. These identifiers serve as ways to refer to the datasets in a time and/or version unaware fashion (i.e. diachronic citations). On the other hand, dataset instantiations define temporal versions of diachronic datasets, holding information on how and when a particular dataset was relevant and actively used.

**Definition 1**. A diachronic dataset $\mathcal{D}$ is defined as a set $\{d, m\}$ where $d$ is a set of dataset versions $\{d_1, \ldots, d_n\}$ and $m$ is a collection of metadata annotations associated with $\mathcal{D}$. Diachronic datasets usually carry housekeeping information about creation, modification etc. in the archiving context, which is included in $m$. In **Figure 4**, *ex:EFO* represents a diachronic dataset that describes the EFO ontology through time. The same example entity can be seen in **Table 1** in an example RDF serialization.

**Definition 2.** A dataset version, or instantiation, $d$ is defined as a set $\{\mathcal{R}, \mathcal{S}, t, m\}$ where $\mathcal{R}$ is a record set and $\mathcal{S}$ is a schema set, while $t$ is a collection of temporal information associated with $d$, and $m$ is a collection of non-temporal annotations associated with $d$. In **Figure 4**, instantiations of *ex:EFO* can be seen as versions 2.35 and 2.36. These can also be seen in **Table 1** in their serialized form.

*Record sets and Schema Sets*. Record sets are collections of data entries (e.g. tuples, triples) over a given subject/primary key within a particular dataset instantiation.

Given a record set and the dataset's metadata information, the dataset instantiation can be queried and reproduced in its original form. Similarly, schema set contains all schema-related entities (e.g. table definitions in the relational case, ontology entities in the ontological case etc.). Keeping data objects separate from schema objects makes versions interpretable by different schemata (e.g. new schema on old data or vice versa).

**Definition 3**. A record set $\mathcal{R}$ is defined as a set $\{r, m\}$, where $r$ is a set of records $\{r_1, .., r_n\}$ and $m$ is a collection of associated metadata for $\mathcal{R}$. A record set $\mathcal{R}$ is always enclosed in the scope of a dataset instantiation $d$, as discussed in Definition 2. The record set for version 2.35 of the EFO ontology can be seen in **Figure 4** and **Table 1** as *ex:recordSet_2.35*.

**Definition 4**. A schema set $\mathcal{S}$ is defined as a set $\{e, m\}$, where $e$ is a set of schema objects $\{e_1, .., e_n\}$ and $m$ is a collection of associated metadata for $\mathcal{S}$. A schema set $\mathcal{S}$ is always enclosed in the scope of a dataset instantiation $d$, as discussed in Definition 2. The schema set for version 2.35 of the EFO ontology can be seen in **Figure 4** as *ex:schemaSet_2.35*.

*Data and Schema Objects*. Data objects consist of *records* and *record attributes*. A record represents a most granular data entry about a particular evolving entity. Records are uniquely identified in order to make record-level annotation feasible in order to attribute provenance, temporality and changes on them. A record serves as a container of one or more record attributes. Every data record is broken down to assertions (facts) that can be expressed as RDF triples. In this sense, a record reifies the predicate-object pairs for a fixed subject. These predicate-object pairs are called record attributes. For instance, a tuple from a relational table is considered to be a record describing the tuple's primary key, with each relational attribute being a record attribute. In [17,18] we describe in details how data records from relational, multidimensional and RDF models can be mapped to data objects in our model. Schema objects represent the schema-related entities of the archived datasets given the dataset's source model. For instance, the classes along with their class restrictions of an ontology, the properties and their definitions (domains, ranges, meta properties depending on the expressivity) are modelled as schema objects. Similarly to data objects, the goal is to provide a reusable modelling mechanism for identifying and referring to schema elements and their evolution across datasets. In this way, schema evolution is captured by annotating schema elements with schema changes.

**Definition 5**. A record $r$ is defined as a set $\{s, a, m\}$ where $s$ is the identifier, or subject, of $r$, $a$ is a collection of record attributes $\{a_1, ..,. a_n\}$ and $m$ is a collection of associated metadata for $r$. In **Figure 4**, an example record can be seen as a part of *ex:recordSet_2.35*. A record describing the experimental factor *EFO_0000887* can be seen in **Table 1**.

**Definition 6**. A record attribute $a$ is defined as a set $\{p, o, m\}$ where $p$ and $o$ are predicate-object pairs and $m$ is a collection of metadata associated with $a$. In **Figure 4**, the record attributes for version 2.35 are the direct children of the aforementioned record. In **Table 1**, two record attributes that describe the label of *EFO_0000887* are shown, with the use of the *rdfs:label* property.

*Diachronic Resources and resource instantiations*. Similarly to diachronic datasets, a diachronic resource represents a time-agnostic information entity. The resource instantiation captures the resource evolution across time and its realization over a versioned dataset's records. The definition of a resource consists of two parts; the resource identification definition gives the way an instantiated resource is identified within the archive. The resource description definition provides the way a resource is evaluated over the records of a particular dataset instantiation. Resources can be versatile in nature across datasets and data formats. For example, given an ontology and its instantiation, each class instance can describe a resource identified by the respective URI. Given a table of employees in a relational database, a resource in this sense can be a particular employee identified by his primary key. Finally, in a multidimensional dataset, a resource can be a specific observation identified by the values of the constituent dimensions.  More complex definitions of resources are allowed and, in fact, encouraged for capturing more high-level, curator specific semantics of evolution and dataset dynamics.

**Definition 7**. A diachronic resource $\mathcal{E}$ is defined as a set $\{\mathcal{I}, q, m\}$ where $\mathcal{I}$ is a set of resource instantiations $\{\mathcal{I}_1, \ldots, \mathcal{I}_n\}$, $q$ is a description definition and $m$ is a collection of metadata associated with $\mathcal{E}$. The description definition $q$ is a DIACHRON query.

**Definition 8**. A resource instantiation $\mathcal{I}$ is defined as a set $\{g, t, m\}$ where g is a set of data records $\{r_1, .., r_n\}$, $t$ is the temporal information associated with $\mathcal{I}$ and m is a collection of metadata associated with resource $\mathcal{I}$.

*Change sets*. Changes come in Change Sets between two dataset instantiations of a diachronic dataset. These are comprised of changes between record sets, changes between schemata and changes between resource instantiations of the two datasets under comparison.

**Definition 9**. A change set $\mathcal{C}$ is defined as a set $\{c, m\}$ where $c$ is a set of changes $\{c_1, \ldots, c_n\}$ and $m$ is a collection of metadata associated with $\mathcal{C}$. The change set between versions 2.35 and 2.36 of the EFO ontology can be seen in **Figure 4** as *ex:changeSet_2.35-2.36*. Furthermore, the same change set can be seen in **Table 1** in a serialized form.

The proposed data model provides a conceptual way of uniformly representing low-level and high-level evolving entities. Within the context of our model, an evolving entity is a dataset instantiation (affected by changes in its schema and contents), a schema object, a data object or finally a resource instantiation object. This gives us a uniform way to model evolution and annotate entities at different levels of granularities with information related to the changes affecting them. Furthermore, it enables us to enrich evolving entities with metadata related to the way these entities are published on remote sites and collected in the archive, such as provenance information, quality and trust.

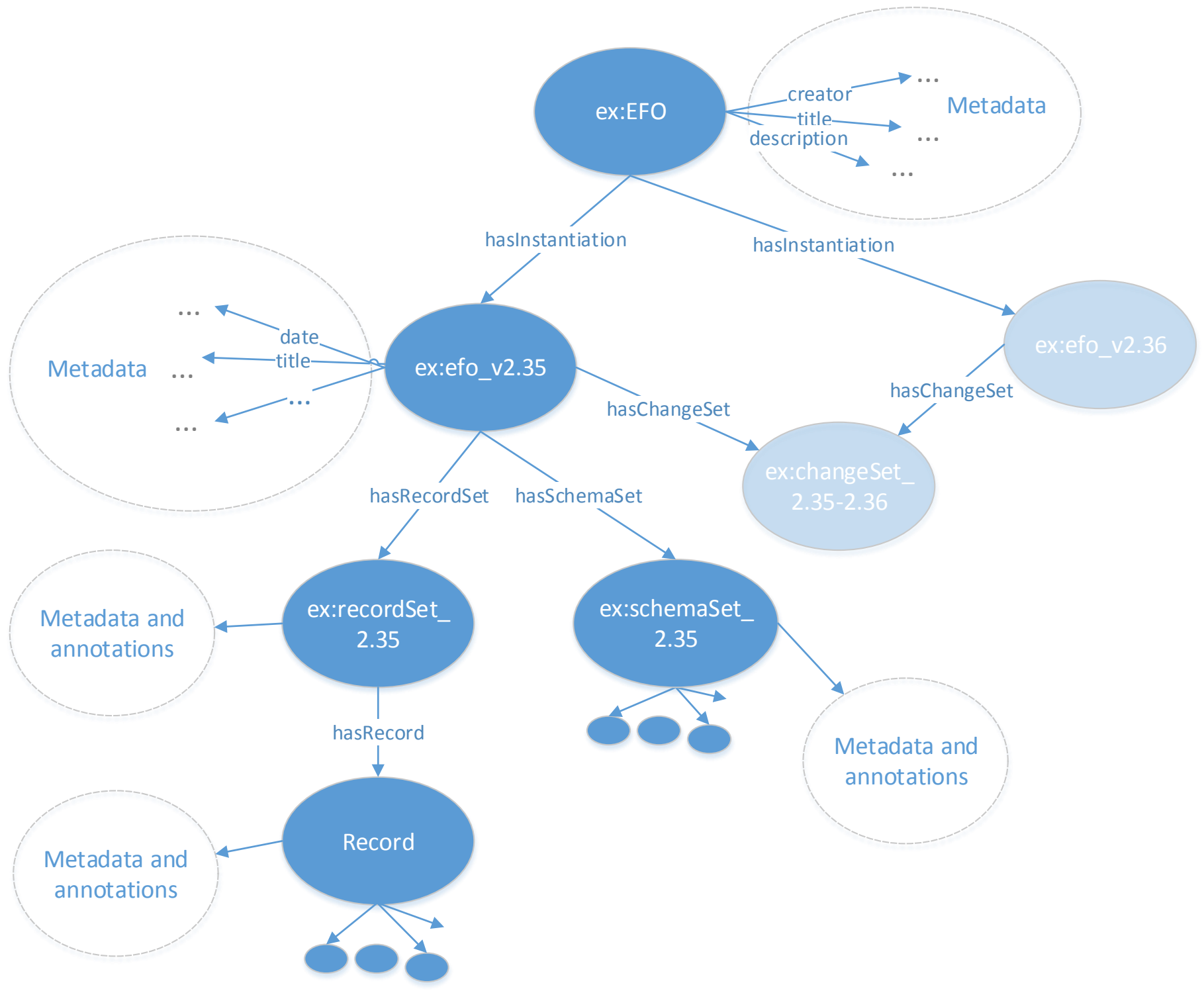

**Figure 4.** An example of a diachronic dataset (ex:EFO) that has two dataset instantiations (versions 2.35 and 2.36). The record and schema sets of version 2.35 can be seen in bold blue, while version 2.36 and the change set that is shared between 2.35 and 2.36 can be seen in pale blue.

**Table 1.** Example RDF serialization of a diachronic dataset ex:EFO, two dataset instantiations (versions) ex:EFO_v.235 and ex:EFO_v.236, in their respective record sets ex:RecordSet_2.35 and ex:RecordSet_2.36. The two record sets contain one record about efo:EFO_0000887, an original instance of the EFO ontology, which shows how its label changes its capitalization between versions. Note that the prefix “ex” is an example prefix. A change set containing a sample *LabelModificationChange* can also be seen.

```
ex:EFO          rdf:type diachron:DiachronicDataset ;
                dcterms:creator "European Bioinformatics Institute" ;
                diachron:hasInstantiation ex:EFO_v2.35 ;
                diachron:hasInstantiation ex:EFO_v2.36 ;
                diachron:hasChangeSet ex:ChangeSet_2.35_2.36 .
ex:EFO_v2.35 rdf:type diachron:Dataset ;
                dcterms:date "2015-01-02"^^xsd:date ;
                diachron:hasRecordSet ex:RecordSet_2.35.
ex:EFO_v2.36 rdf:type diachron:Dataset ;
                dcterms:date "2015-02-02"^^xsd:date ;
                diachron:hasRecordSet ex:RecordSet_2.36.
ex:RecordSet_2.35 rdf:type diachron:RecordSet ;
                diachron:hasRecord ex:Record_1
ex:Record_1 diachron:subject efo:EFO_0000887 ;
                diachron:recordAttribue ex:RecordAttribute_1 .
ex:RecordAttribute_1 diachron:predicate rdfs:label ;
                diachron:object "liver"
ex:RecordSet_2.36 rdf:type diachron:RecordSet ;
                diachron:hasRecord ex:Record_2 .
ex:Record_2 diachron:subject efo:EFO_0000887 ;
                diachron:recordAttribue ex:RecordAttribute_2 .
ex:RecordAttribute_2 diachron:predicate rdfs:label ;
                diachron:object "LIVER" .
ex:ChangeSet_2.35-2.36 rdf:type diachron:ChangeSet ;
                diachron:oldVersion ex:EFO_v2.35 ;
                diachron:newVersion ex:EFO_v2.36 ;
                diachron:hasChange ex:Change1 .
ex:Change1 rdf:type diachron:LabelModificationChange ;
                diachron:parameter1 ex:RecordAttibute_1 ;
                diachron:parameter2 ex:RecordAttibute_2 .
```

# 4 The DIACHRON Query Language

## 4.1 Requirements and Overview

The DIACHRON model provides metadata placeholders in different granularities, from the dataset to the record level. In this section, we motivate the need for an appropriate query language that exploits the specificities of the data model and provides ways to achieve the following:

- *Dataset and version listing*: Retrieve lists of datasets stored in the archive, as well as lists of the available versions of a given dataset. These can either be exhaustive or filtered based on temporal, provenance or other metadata criteria.
- *Data queries*: Retrieve part(s) of a dataset that match certain criteria.
- *Longitudinal queries*: As above but with the timeline of all types of diachronic entities. Temporal criteria can be applied to limit the timeline (specific versions or time periods), or successive versions.
- *Queries on Changes*: Retrieve changes between two concurrent versions of an entity (dataset, resource etc.). Limit results for specific type of changes, or for a specific part of the data.
- *Mixed Queries on Changes and Data*: Retrieve datasets or parts of datasets that are affected by specific types of changes.

In this section, we propose the DIACHRON Query Language (DIACHRON QL), to tackle these requirements, and we discuss its design and implementation as an extension of SPARQL. The basis of the query language is the DIACHRON graph pattern, which, in the context of extending SPARQL, is a specialization of a SPARQL graph pattern, thus making SPARQL queries valid DIACHRON QL queries. New keywords are defined in order to cover the model's characteristics and allow the user to query archived data intuitively, without the need to know the specificities of the implementation. In plain SPARQL engines, or any other query engine basis, the user would need to know how the DIACHRON model is implemented in the system, and how its entities are mapped to the system's underlying information retrieval engine. With the use of a dedicated query language, we abstract the implementation details to the DIACHRON QL syntax. DIACHRON QL introduces keywords that allow defining the scope of a query with respect to the matched diachronic datasets and their versions, their change sets, or both.

### 4.2 DIACHRON QL basics

Given the above, diachronic datasets, versions and change sets can be bound to variables with the use of `DATASET` or `CHANGES`. This is simply done by using variables instead of explicit URIs, inside the query body, i.e. not in a `FROM` clause. For example, consider the case where we want to retrieve all the information (predicate-object pairs) associated with the protein *efo:EFO_0004626*, and find out what the state of this information is for all the dataset versions of the EFO ontology it appears in (and what are those versions). That is, the dataset versions as well as the actual information are to be retrieved. In DIACHRON QL this can be written as follows:

```
SELECT ?version ?p ?o WHERE {
     DATASET <EFO> AT VERSION ?version {
            efo:EFO_0004626  ?p  ?o
     }
```

```
}
```

This will retrieve all versions of EFO joined with predicate-object pairs for the protein *efo:EFO_0004626*. If we want to retrieve the records these predicate-object pairs appear in, without querying for the particular dataset versions. We can retrieve the URIs of the DIACHRON records these triples appear in by modifying the query as follows:

```
SELECT ?rec ?p ?o FROM DATASET <EFO> WHERE {
          RECORD ?rec {efo:EFO_0004626  ?p  ?o}
}
```

With the optional use of the `RECATT` keyword we can retrieve the URIs of the record attributes of a matched record. The previous query would become:

```
SELECT ?rec ?ra ?p ?o FROM DATASET <EFO> WHERE {
          RECORD ?rec {efo:EFO_0004626
                RECATT ?ra {?p  ?o}
          }
}
```

When writing a DIACHRON graph pattern, the query can either contain simple triple patterns, or more verbose constructs that take into account the archive data model and structure. Specifically, the simple triples will match the de-reified data, whereas the `RECORD` and `RECATT` (abbreviation of '*record attribute*') blocks will also take into account a triple's record or record attribute.

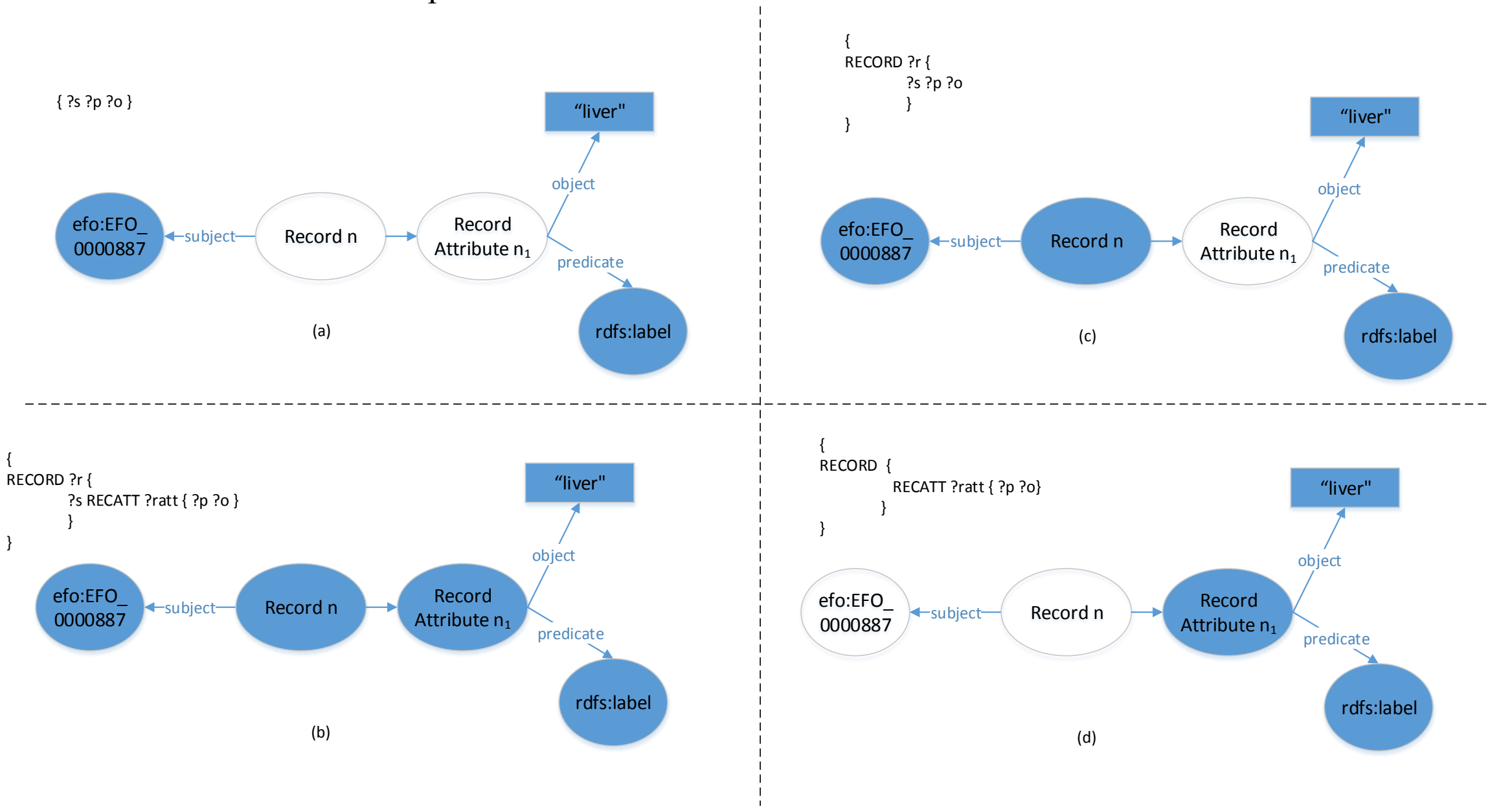

**Figure 5**. (a) matches in a simple triple query, (b) matches a blown-out version of the same query with the RECORD and RECATT terms, selecting both data and structural elements. (c) matches subject, predicate, object and record, (d) matches predicate, object and record attribute.

This is further exemplified in Figure 5 where we show how term and variable use is reflected on the matched graph of a particular reified triple. This way, metadata (e.g. temporal, provenance) of the records and/or record attributes can be queried as well as combined with data queries. It should be noted that in the simplest case where only the data are of interest, the query does not need to include `RECORD` or `RECATT` blocks.

### 3.2 Query Syntax and Examples

DIACHRON QL clauses are formally described in the following section and an overview of them is presented in **Table 2**. In **Table 3** usage examples are presented for all DIACHRON QL clauses.

*`FROM DATASET <diachronicURI> [[AT VERSION <instantiationURI>]]`*

The `FROM DATASET` keyword is followed by a URI of a diachronic dataset to declare the dataset scope of the query. If no `FROM DATASET` is given, then the whole corpus of datasets is queried. The optional `AT VERSION` keyword limits the selected diachronic dataset to a specific dataset instantiation. No variables can be given in any of the parameters of `FROM DATASET AT VERSION`.

**Table 2. The DIACHRON query language syntax in E-BNF.**

| | |
|---|---|
| *DiachronQuery* := | ‘DIACHRON‘<br>‘SELECT’ (‘DISTINCT‘)? (Var+\|’*’)<br>*Source_Clause**<br>‘WHERE‘ *Where_Clause** |
| *Source_Clause* := | ( ‘FROM DATASET’ <*URI*> [‘AT VERSION’ <*URI*>] \|<br>‘FROM CHANGES’ <*URI*> [‘BEFORE VERSION’ <*URI*> \|<br>‘AFTER VERSION’ <*URI*> \| ‘BETWEEN VERSIONS’ <*URI*>+2] ) |
| *Where_Clause* := | ( *Diachron_Pattern*<br>*[‘UNION’ Diachron_Pattern]*<br>*[‘OPTIONAL‘ Diachron_Pattern] )* |
| *Diachron_Pattern* := | (*Source_Pattern Basic_Archive_Graph_Pattern* ) |
| *Source_Pattern* := | ((‘DATASET‘ <VarOrURI> [‘AT VERSION’ <VarOrURI>]) \|<br>(‘CHANGES’ <VarOrURI> [‘BEFORE VERSION’ <VarOrURI>]) \|<br>(‘CHANGES’ <VarOrURI> [‘AFTER VERSION’ <VarOrURI>]) \|<br>(‘CHANGES’ <VarOrURI> [‘BETWEEN VERSIONS’ <VarOrURI>+2])) |
| *Basic_Archive_Graph_Pattern* := | ‘{‘ *SPARQL_Triples_Block* Record_Block* Change_Block** ‘}’ |
| *Record_Block* := | ‘RECORD‘ <VarOrURI> ‘{‘ |

<table>
<tr><td></td><td><VarOrURI> ((<VarOrURI>+2 ‘.’)*) |<br>(‘RECATT‘ <VarOrURI> ‘{‘ <VarOrURI>+2 ‘}’)*<br>‘}’</td></tr>
<tr><td>Change_Block :=</td><td>‘CHANGE‘ <VarOrURI> ‘{‘<br>(<VarOrURI>+2 ‘.’)*<br>‘}’</td></tr>
<tr><td>SPARQL_ Triples_Block :=</td><td>As defined in the SPARQL recommendation[5].</td></tr>
</table>

*FROM CHANGES <diachronicURI> [[BETWEEN VERSIONS <version1URI> <version2URI>] || [BEFORE VERSION <versionURI>] || [AFTER VERSION <versionURI>]]*

FROM CHANGES is used to query change-sets directly. Optionally, it is immediately followed by a URI of a diachronic dataset that defines the diachronic dataset to be queried on its changes. If no URI is given, then all existing change sets will be used to match the query body. FROM CHANGES can optionally be used with BETWEEN VERSIONS, BEFORE or AFTER VERSION to limit the scope of the changes.

*DATASET <URI | ?var> [[AT VERSION <URI | ?var>]] { (query) }*

The DATASET keyword differs from FROM DATASET in that it is found inside a query body. It is followed by a URI/variable of a diachronic dataset to declare or bind the scope of the graph. DATASET is inside a WHERE statement and is followed by a graph pattern, on which the dataset restriction is applied. It is optional, meaning that if no DATASET is given, then the whole corpus of datasets will be queried, or the datasets defined in the FROM DATASET clause. The AT VERSION keyword, when applied to a DATASET statement inside a WHERE clause, is used to either define a specific dataset instantiation or bind dataset instantiations to a variable for the graph pattern that follows. However, AT VERSION is optional and if no specific dataset instantiation URI or variable is declared, AT VERSION is omitted. An example of matching both triples and versions can be seen in **Figure 6**.

5 http://www.w3.org/TR/sparql11-query/

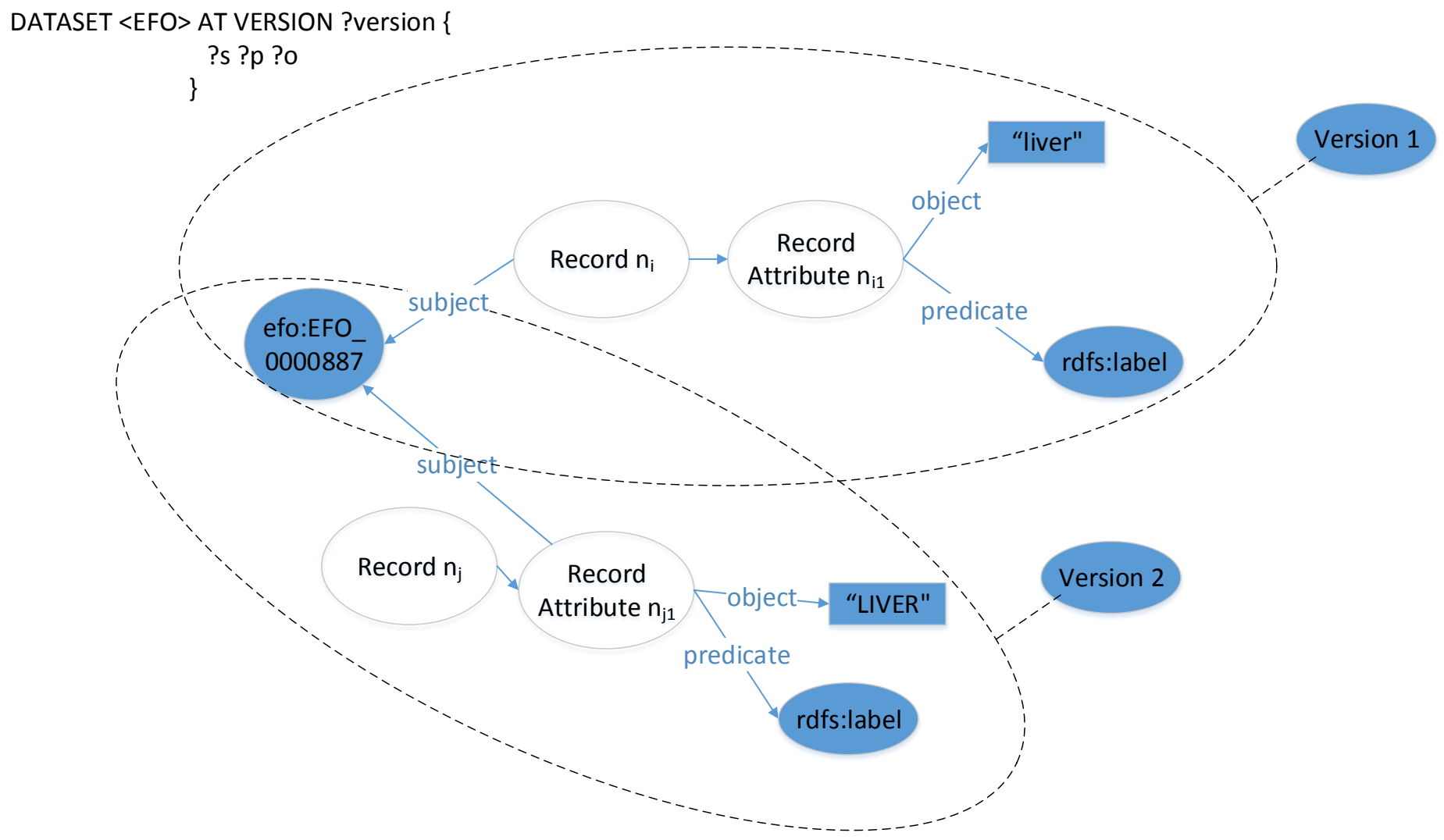

**Figure 6**. Matching a reified triple in a query with variable versions. Blue nodes are selected by the query.

*RECORD <record_URI | ?record_var>*
*{<subjectURI | ?subject_var > ATTRIBUTE_pattern}*

`RECORD` is used inside the body of a graph pattern for querying either a specific DIACHRON record or to match DIACHRON records in the pattern. It is followed by a record URI/variable. If neither of those is declared, the `RECORD` keyword can be omitted. Following `RECORD` is a block containing a graph pattern that can either be of SPARQL form, or used in conjunction with the `RECATT` keyword.

**Table 3. Query language keywords and usage examples.**

| Keyword | Parameters | Usage example |
|---|---|---|
| **SELECT** | variable list | SELECT ?x, ?y, ?z |
| **FROM DATASET** | URI of diachronic dataset | SELECT ?x, ?y, ?z<br>FROM DATASET <efo-protein-sample> |
| **FROM DATASET AT VERSION** | URI of dataset instantiation | SELECT ?x, ?y, ?z<br>FROM DATASET <efo-protein-sample> AT VERSION <v1> |
| **FROM CHANGES** | URI of diachronic dataset | SELECT ?x, ?y, ?z<br>FROM CHANGES <efo-protein-sample> |
| **FROM CHANGES ... BETWEEN VERSIONS** (params) | URIs of dataset instantiations to define the change scope | SELECT ?x, ?y, ?z<br>FROM CHANGES <efo-protein-sample> BETWEEN VERSIONS <$v_m$>, <$v_n$> |
| FROM CHANGES … **AFTER / BEFORE VERSION** (params) | URI of dataset instantiation to define the start/end of the change scope | SELECT ?x, ?y, ?z<br>FROM CHANGES <efo-protein-sample> AFTER / BEFORE VERSION <$v_m$> |
| **WHERE** { (params) } | DIACHRON patterns | SELECT ?x, ?y, ?z<br>FROM DATASET <efo-protein-sample><br>WHERE {<br>?x a efo:Protein ; ?y ?z .<br>} |
| **DATASET** (params) | URI or variable of diachronic dataset | SELECT ?x, ?y<br>WHERE {<br>DATASET ?x {<br>?s a efo:Protein.<br>}<br>DATASET ?y {<br>?s dcterms:creator "EBI"<br>}<br>} |
| **DATASET** … **AT VERSION** (params) | URI or variable of dataset instantiation | SELECT ?x, ?y<br>WHERE {<br>DATASET ?x AT VERSION ?var {<br>?s a efo:Protein.<br>}<br>DATASET ?y AT VERSION <v1> {<br>?s dcterms:creator "EBI"<br>}<br>} |
| **RECORD** (params) | URI or variable of DIACHRON record | SELECT ?x, ?r, ?y<br>WHERE {<br>DATASET ?x AT VERSION ?var {<br>RECORD ?r {?s a efo:Protein}<br>}<br>DATASET ?y AT VERSION <v1> {<br>?s dcterms:creator "EBI"<br>}<br>} |
| **RECATT** (params) | URI or variable of a DIACHRON record attribute | SELECT ?var, ?r, ?ra<br>WHERE {<br>DATASET <efo> AT VERSION ?var {<br>RECORD ?r {<br>?s RECATT ?ra {rdf:type efo:Protein}<br>}<br>}<br>} |
| **CHANGES** (params) | URI of diachronic dataset or variable | SELECT ?c, ?param1, ?value1<br>WHERE {<br>CHANGE ?c {?param1 ?value1 }<br>} |
| **CHANGES** ... **BETWEEN VERSIONS** (params) | URIs of dataset instantiations or variables to define the change scope | SELECT ?v1, ?v2, ?c<br>WHERE {<br>CHANGES <EFO> BETWEEN VERSIONS ?v1, ?v2 {<br>?c rdf:type co:Add_Definition ;<br>?p1 [co:param_value ?o3 . rdf:type co:ad_n1 ] ;<br>?p2 [co:param_value ?o4 . rdf:type co:ad_n2 ]<br>}<br>} |
| **CHANGES** … **AFTER / BEFORE VERSION** (params) | URI of dataset instantiation or variable to define the start/end of the change scope | SELECT ?s ?p ?o<br>WHERE {<br>CHANGES <efo-protein-sample> BEFORE/AFTER VERSION <$v_m$> { ?s ?p ?o}<br>}<br>} |
| **CHANGE** (params) | URI of change or variable | SELECT ?v1, ?v2, ?c, ?p ?o WHERE {<br>CHANGES <EFO> BETWEEN VERSIONS ?v1 ?v2{<br>CHANGE ?c {?p ?o}<br>} |

*RECATT <recattURI | ?recatt_var>*
*{ <predicateURI | ?predicate_var> <objectURI | ?var> }*

`RECATT` is used inside a `RECORD` block and separates the subject of a DIACHRON record with the record attributes that describe it. It is followed by a URI/variable. If no specific record attribute needs to be queried or matched in a variable, `RECATT` can be omitted.

*CHANGES <diachronicURI | var> [[BETWEEN VERSIONS <version1URI | ?var1>] || [BEFORE VERSION <versionURI | var1>] || [> AFTER VERSION <versionURI | var1>]]*

`CHANGES` is used to limit the scope of a block within a larger query into a particular change set, or match change sets to a variable. If no URI is given, then all existing change sets will be used to match the query body. CHANGES can optionally be used with `BETWEEN VERSIONS`, `BEFORE VERSION` or `AFTER VERSION` to limit the scope of the changes or bind the dataset versions that match the change set pattern to variables.

*CHANGE <changeURI | ?change_var>*

The `CHANGE` keyword is used to query a particular change in a fixed query block within a larger query pattern. It is followed by a specific change URI or a variable to be bound. The succeeding block is used to declare the change parameters in a predicate-object manner.

### 4.3 DIACHRON QL formal definitions

In order to formally describe DIACHRON QL as a SPARQL extension, it is necessary to address the DIACHRON model as an extension of RDF, in a manner similar to [11,14,22,23]. Let *I, B, L, V* be infinite, pairwise disjoint sets of IRIs, blank nodes, literals and variables respectively. An RDF triple $t$ is a triple $(\mathrm{s}, \mathrm{p}, \mathrm{o}) \in (I \cup B) \times (I) \times (I \cup B \cup L)$, where $s$ is the subject, $p$ is the predicate and $o$ is the object of the triple. An RDF graph is a collection of triples g = $\{t_1, t_2, \ldots, t_n\}$. The union $(I \cup B \cup L)$ is denoted as $T$ and represents all possible bound values any node in an RDF graph can take. The set of all RDF graphs is denoted as $G$. Given the above, we define the DIACHRON model entities as follows:

Definition 10. A record attribute $a$ is a tuple $(t, g)$ where t is an RDF triple, and g is a metadata subgraph for $a$. In essence a record attribute associates an RDF triple $t$ with its metadata, expressed as an RDF graph $g$. We denote as $G_a \subseteq G$ the set of all record attributes, and as $I_a \subseteq I$ the set of all record attribute IRI nodes.

Definition 11. A record r is defined as a tuple $(G_a^s, g)$, where $G_a^s \subseteq G_a$ is a set of record attributes over subject s, and g is a metadata subgraph associated with r. The set $G_a^s$ is only relevant to the particular context and is not meant to be an exhaustive list of triples with s as common subject. We denote as $G_r \subseteq G$ the set of all records, and as $I_r \subseteq I$ the set of all record IRI nodes.

Definition 12. A record set $R$ is defined as a tuple $(G_r', g)$, where $G_r' \subseteq G_r$ is a set of records, and g is a metadata subgraph associated with $R$. We denote as $G_R \subseteq G$ the set of all record sets, and as $I_R \subseteq I$ the set of all record set IRI nodes.

Definition 13. A schema set $S$ is defined as a tuple $(G_S', g)$, where $G_S' \subseteq G_r$ is a set of schema elements, and g is a metadata subgraph associated with $S$. We denote as $G_S \subseteq G$ the set of all schema sets, and as $I_S \subseteq I$ the set of all schema set IRI nodes.

Definition 14. A dataset instantiation d is a tuple $(G_R', G_S', g)$ where $G_R' \subseteq G_R$ and $G_S' \subseteq G_S$ are the record set and schema set of the instantiation. We denote as $G_d \subseteq G$ the set of all dataset instantiations, and as $I_d \subseteq I$ the set of all dataset instantiation IRI nodes.

Definition 15. A diachronic dataset D is a tuple $(G_d', g)$ where $G_d' \subseteq G_d$ is an arbitrary set of dataset instantiations as per Definition 14. We denote as $G_D \subseteq G$ the set of all diachronic datasets, and as $I_D \subseteq I$ the set of all diachronic dataset IRI nodes. Similarly to SPARQL we allow for blank nodes and literals to be identifier values, as well as triple subjects, even though in practice this is not supported by most frameworks. Note further that the metadata subgraph can be an empty graph. This allows for definitions of datasets and other DIACHRON entities without necessarily associating metadata with them.

The above definitions serve to regard the entities of the DIACHRON model as extensions of RDF. Examples of these are shown in **Figure 4** and **Table 1**, as discussed in Section 3.

In order to define the syntax of DIACHRON QL, we briefly recall the notion of a SPARQL graph pattern presented in [22]. A SPARQL graph pattern expression is defined recursively as follows:

1. A tuple from $(T \cup V) \times (I \cup V) \times (T \cup V)$ is a graph pattern.
2. If $P_1$ and $P_2$ are graph patterns, then ($P_1$ AND $P_2$), ($P_1$ OPT $P_2$), and ($P_1$ UNION $P_2$) are graph patterns.
3. If P is a graph pattern and R is a SPARQL built-in condition, then the expression (P FILTER R) is a graph pattern.

Given this, a DIACHRON QL graph pattern expression (DGP) is defined hierarchically and recursively as follows:

1. A SPARQL graph pattern P is a DGP.
2. If $X \in (I_a \cup V)$ then (X RECATT P) is a DGP (a record attribute pattern).
3. If $X \in (I_r \cup V)$ then (X RECORD P) is a DGP (a record pattern).
4. If $P$ is a DGP, $X \in (I_D \cup V)$ and $Y, Z \in (I_d \cup V)$ then :
   (a) (( (DATASET X) AT VERSION Y ) P),
   (b) (( (DATASET X) AFTER VERSION Y ) P),
   (c) (( (DATASET X) BEFORE VERSION Y ) P),
   (d) (( (DATASET X) AFTER VERSIONS Y, Z ) P)
   (e) ((DATASET X) P)

   are DGPs (dataset instantiation patterns).
5. If $P_1$ and $P_2$ are DGPs, then the following are DGPs:
   (a) $P_1\ AND\ P_2$
   (b) $P_1\ OPT\ P_2$
   (c) $P_1\ UNION\ P_2$

DIACHRON QL built-in conditions for filtering are similar to [22] and are not further addressed in this paper. Furthermore, formal definitions for CHANGES are similar to DATASET and are not further discussed in the name of readability. Examples on all keywords and constructs of DIACHRON QL can be seen in **Table 3**.

**Semantics of DIACHRON QL graph pattern expressions**

We are now ready to define the semantics of DPG expressions. Borrowing the notation of [22], a SPARQL mapping, or substitution, μ is defined as a partial function $\mu : V \rightarrow T$ for a subset $V' \subseteq V$, such that the variables in $V'$ are replaced with values from T as is defined in μ. The domain of μ: $dom(\mu)$ is the subset of V where μ is defined. A pair of mappings $\mu_1$ and $\mu_2$ exhibits compatibility when for all $v \in dom(\mu_1) \cap dom(\mu_2)$, it holds that $\mu_1(v) = \mu_2(v)$. Let $\Omega_1$ and $\Omega_2$ be sets of mappings, then the join, the union, and the difference between $\Omega_1$ and $\Omega_2$ are defined as follows:

- $\Omega_1 \bowtie \Omega_2 = \{\mu_1 \cup \mu_2 \mid \mu_1 \in \Omega_1, \mu_2 \in \Omega_2$ are compatible$\}$,
- $\Omega_1 \cup \Omega_2 = \{\mu \mid \mu \in \Omega_1$ or $\mu \in \Omega_2\}$,
- $\Omega_1 \setminus \Omega_2 = \{\mu \in \Omega_1 \mid$ for all $\mu' \in \Omega_2$, $\mu$ and $\mu'$ are not compatible$\}$.

Finally, the left outer-join (OPTIONAL) is defined as:

- $\Omega_1 ⟕ \Omega_2 = (\Omega_1 \bowtie \Omega_2) \cup (\Omega_1 \setminus \Omega_2)$.

Given the above, the notion of mapping remains the same in DIACHRON QL.

In DIACHRON QL, the hierarchical relationship between entities enables graph patterns to be limited in *scopes*, with respect to the DIACHRON model. Evaluating a triple pattern within the scope of two different record patterns can result in different output, and also enables pattern expressions involving the binding of DIACHRON model entities as well.

Formally, we need to define what the scope of a graph pattern is. A scope $\sigma$ is a function $\sigma: P \rightarrow T' \subseteq T$ that maps a graph pattern P to a closed set $T'$, so that any mapping $\mu_p$ of P is only valid with respect to $T'$, i.e. $\mu_P \subseteq \sigma(P)$. Given this, we go on to define the *lowest wrapping scope* $\lambda$ as a partial function $\lambda : P \rightarrow \sigma(I \cup V)$ that maps P with a scope, such that the variables in P are mapped to elements in that scope, and there exists no other scope that is a subset of the one derived from λ. This implies that any graph pattern P is equipped with a function $\lambda(P) \in \sigma(I \cup V)$ such that $\mu(P) \in \lambda(P)$ and $\nexists \lambda'(P) \neq \lambda(P) | \lambda'(P) \subseteq \lambda(P)$. Furthermore, we denote with $\lambda_{D'}(P)$ when $\lambda(P)$is limited to a specific subset of diachronic datasets and dataset instantiations $D'$. Intuitively, a lowest wrapping scope for a particular query is the lowest entity type in the DIACHRON model hierarchy where P is expressed. For example, a record attribute pattern $P_a$ in a query is nested within a record pattern $P_r$ and a dataset instantiation pattern $P_d$. Then $\lambda(P_a) = \sigma(P_r)$ and $\lambda(P_r) = \sigma(P_d)$. An example of scoping in a DIACHRON query can be seen in **Figure 7**.

We are now ready to define the evaluation of a DIACHRON QL graph pattern. Given a diachronic dataset *D* with a set of dataset instantiations *d* over *T*, such that $D' \subseteq D$ is the subset of D in which *d* exists, and DGPs *P*, $P_1$ and $P_2$ defined in $D'$, $D_1'$ and $D_2'$ respectively, then the evaluation of a DGP denoted as $[[\bullet]]_{D'}$ is as follows:

- $[[P]]_{D'} = \{\mu \mid dom(\mu) = var(P) \ and \ \mu(P) \in \lambda_{D'}(P) \}$
- $[[P_1 \ AND \ P_2]]_{D_1',D_2'} = \{\mu = [[P_1]]_{D_1'} \bowtie [[P_2]]_{D_2'} \mid \mu \in \lambda_{D_1'}(P_1) \cap \lambda_{D_2'}(P_2)\}$
- $[[P_1 \ UNION \ P_2]]_{D_1',D_2'} = \{\mu = [[P_1]]_{D_1'} \cup [[P_2]]_{D_2'} \mid \mu \in \lambda_{D_1'}(P_1) \cup \lambda_{D_2'}(P_2)\}$
- $[[P_1 \ OPT \ P_2]]_{D_1',D_2'} = \{\mu = [[P_1]]_{D_1'} ⟕ [[P_2]]_{D_2'} \mid \mu \in \lambda_{D_1'}(P_1) \cup \lambda_{D_2'}(P_2)\}$

Evaluation of filters remains the same as with the original SPARQL specification [22] and is not reported herein. Finally, note that we do not consider the case of named graphs within DIACHRON graph patterns, because the general notion of a SPARQL named graph is specialized in the more refined DIACHRON entity types.

Given a DIACHRON graph pattern expression (( (DATASET X) AT VERSION Y ) P), its evaluation will be equal to the evaluation of P over diachronic dataset X at version Y, i.e. the set of all mappings μ such that $\mu(P) \in \lambda_{X'}(P)$, with $X'$ being the subset of X that contains version Y.

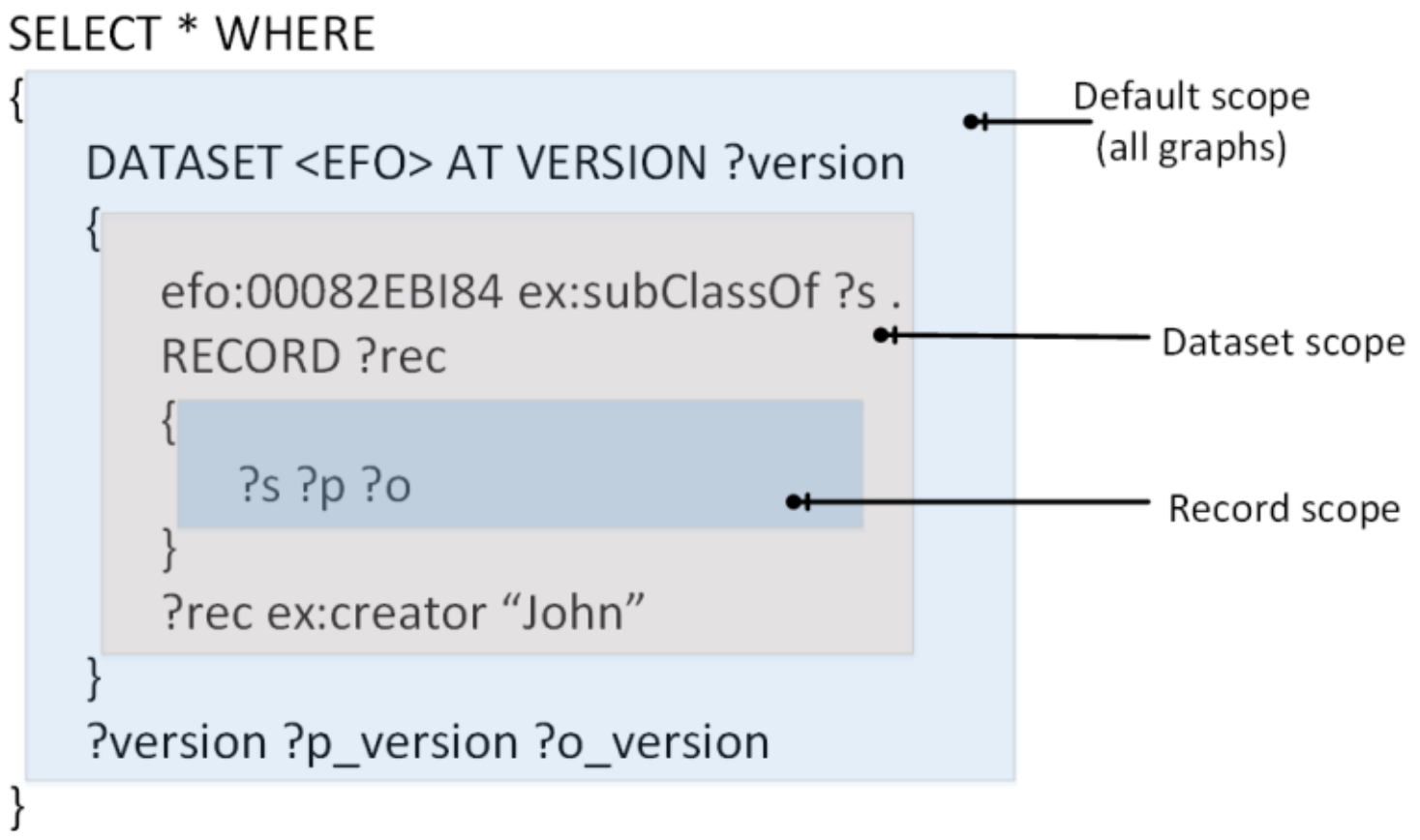

**Figure 7. An example of scopes in a DIACHRON QL query.**

# 5 Implementation

In this section we present the implementation of the proposed query language. We first provide an overview of the overall architecture of the DIACHRON archive. The archive employs the proposed DIACHRON model for storing evolving LOD datasets. The query engine is a core component of the archive, responsible for processing queries expressed in the DIACHRON QL and retrieving data out of the archive.

## 5.1 System architecture

The architecture of the archive and various components of the archive can be seen in Figure 8. The archive's web service interface is exposed via the HTTP protocol as the primary access mechanism of the archive through a RESTful web service API. The Data Access Manager provides low level data management functionality for the archive. It is bound to the specific technology of the underlying store, in our case Openlink Virtuoso 7.1[6], as well as external libraries that provide data access functionality for third-party vendors. For this we used the Jena semantic web framework[7]. It serves as an abstraction layer between the store and the query processor.

[6] http://virtuoso.openlinksw.com/
[7] https://jena.apache.org/

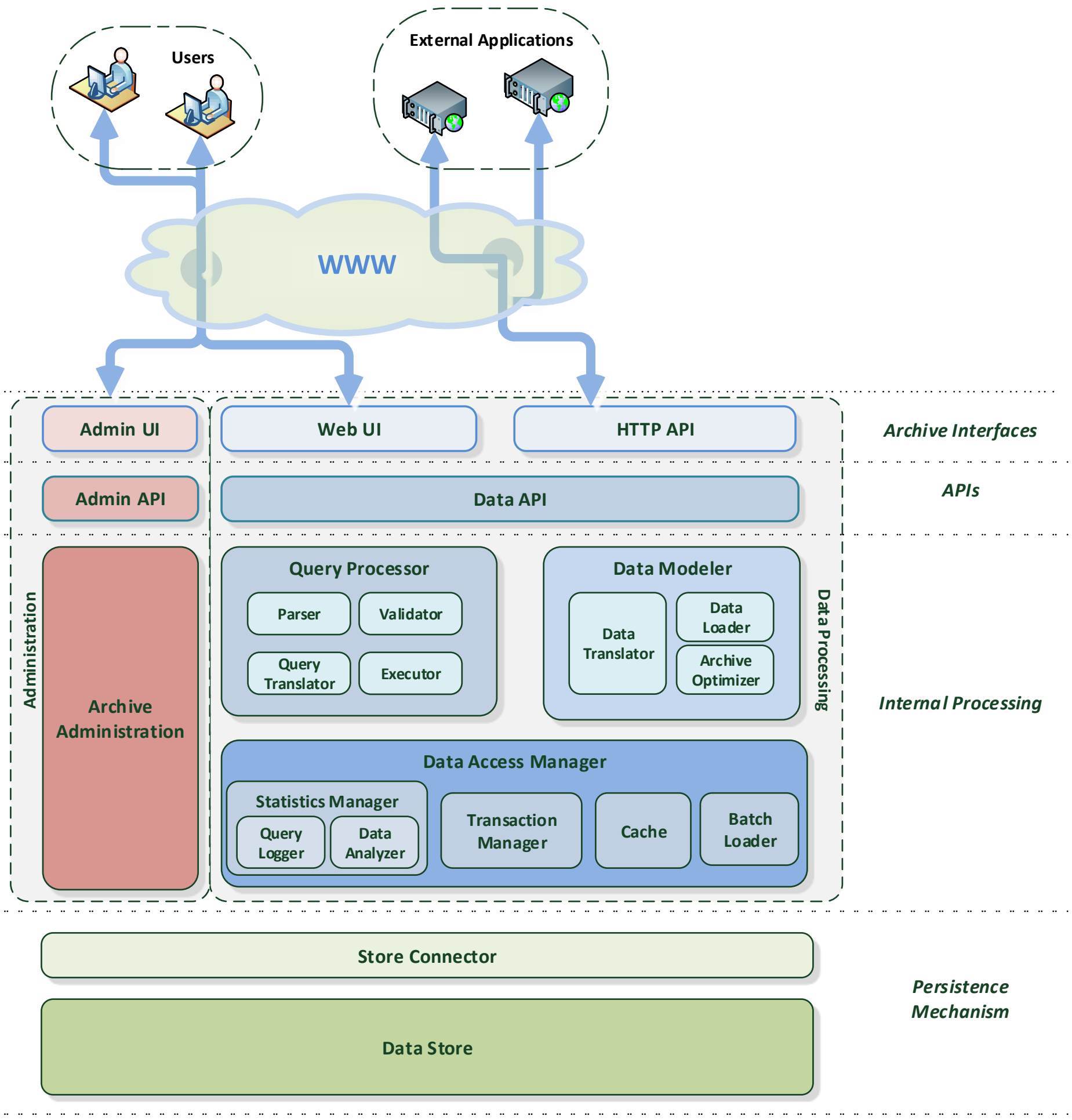

**Figure 8**: Architecture of the archive.

The archive employs a Data Access Manager, a Store Connector, a Data Modeler, an Archive Optimizer and a Query Processor. The Store Connector is the software package that provides an API to other components of the archiving module for communication and data exchange with the underlying store and is implemented with the Virtuoso JDBC Driver package[8]. The Data Store employs a Virtuoso 7.1 instance. The Data Modeler component handles the dataset input functionality and data transformations from the DIACHRON dataset model to the native data model of the store and vice versa, and consists of the Data Translator and the Data Loader. The Archive optimizer component supports the optimization of the datasets' storage method based on various archive strategies as shown in [26] that are not discussed in this paper. It performs analysis of the dataset characteristics and chooses the most efficient storage strategy based on metrics.

[8] http://docs.openlinksw.com/virtuoso/VirtuosoDriverJDBC.html

The Query Processor component is the base mechanism for query processing and thus data access. It consists of the following subcomponents:

- **Validator:** validates the DIACHRON queries for syntactic validity against the DIACHRON QL syntax described in section 3.
- **Query parser:** parses the queries in DIACHRON QL so as to create a structure of elements that correspond to DIACHRON Dataset Entities and DIACHRON query operators.
- **Query Translator:** creates the execution plan of DIACHRON queries by translating the queries in SPARQL. The translator also makes use of the various archive structures implemented in the persistence store and the appropriate indexes and dictionaries. The query translator is the subcomponent that ties the DIACHRON archive module to the specific storage technology of RDF and SPARQL. Translation is further described in section 5.2
- **Executor:** executes the created execution plan step by step and retrieves the raw data from the store so as to build the result set of the query. It uses also the Data Modeler component in order to perform, if necessary, data transformations from the native data model of the underlying store to the DIACHRON dataset model.

### 5.2 Translation of DIACHRON QL to SPARQL

Our implementation is based on mature standards and state of the art triple stores that implement RDF storage and SPARQL querying. This imposes that DIACHRON entities are converted to RDF and queries are mapped to SPARQL expressions. In this context, DIACHRON graph patterns can generally be translated to SPARQL as shown in **Table 4**. However, a direct mapping is not generally possible, as the two models differ conceptually. The actual translation to SPARQL is ultimately dependent on factors that are affected by the implementation at hand, such as the storage policies, the structure of the archive and its dictionary, and the pre-processing requirements of the query engine. For this reason, we have implemented a middle layer between the DIACHRON QL parser and the SPARQL query executor, where the following steps take place:

1. Identification of the query's relevant scope(s), and in-memory mapping to DIACHRON structural elements
2. Extraction and mapping of graph patterns to their respective scopes
3. Conversion of lowest level graph patterns to SPARQL
4. Detection of non-materialized dataset versions that contain possible scope candidates
5. Temporary materialization of non-materialized dataset versions
6. Mapping to final SPARQL query

In the above flow of actions, step 1 is responsible for extracting the scopes $\sigma(P_i)$ for all $P_i$ that are sub-expressions of a DIACHRON query expression P. References to their respective URI nodes or variables point to their subsumed DGP and are stored in memory for future reference. In step 2, we map each scope to its respective DGP found in the query string, and populate the query object in-memory. In step 3, we identify the data-relevant part of the query (i.e. the part that references actual records and attributes), and rewrite it to SPARQL independently of its scope. In step 4, we detect whether a scope is actually materialized in the archive. This step deals with cases where the chosen storage policy differs from full materialization, however it is not in the scope of this work to address the implementation issues of storage policies, the storage-querying trade-off, or storage optimization for contexts with versioning. Furthermore, simple lookups in the dictionary for a given query's scopes is not sufficient to determine which $\sigma$ are eventually referenced, because a scope can be unbound (i.e. a variable). These points are all taken into account in steps 4 and 5. Finally, step 6 relies on the output of the previous steps in order to build one or more SPARQL queries that will be executed by the query engine. Hence, in order to implement DIACHRON QL in a SPARQL setting, the added expressivity of DIACHRON QL over SPARQL is translated to a series of steps, rather than a direct 1:1 mapping of entities and graph pattern expressions.

**Table 4. DIACHRON graph patterns and their translation to SPARQL**

| DIACHRON Pattern (Parsed Syntax) | SPARQL |
|---|---|
| {?s ?p ?o} | { [a evo:Record ;<br>evo:subject ?s ;<br>evo:hasRecordAttribute<br>[ evo:predicate ?p ; evo:object ?o ]]} |
| RECORD ?r {?s ?p ?o} | {?r a evo:Record ;<br>evo:subject ?s ;<br>evo:hasRecordAttribute<br>[evo:predicate ?p ; evo:object ?o]} |
| RECORD ?r {<br>?s RECATT ?ra {?p ?o}<br>} | {?r a evo:Record ;<br>evo:subject ?s ;<br>evo:hasRecordAttribute ?ra .<br>?ra evo:predicate ?p ;<br>evo:object ?o} |
| DATASET <EFO> AT VERSION ?v<br>{<br>RECORD ?r {<br>?s RECATT ?ra {?p ?o}<br>}<br>} | {GRAPH <dataset_dictionary> {<br><EFO> evo:hasInstantiation ?v .<br>?v evo:hasRecordSet ?rs<br>} GRAPH ?rs{<br>?r a evo:Record ;<br>evo:subject ?s ;<br>evo:hasRecordAttribute ?ra .<br>?ra evo:predicate ?p ;<br>evo:object ?o }} |
| FROM DATASET <EFO> AT VERSION <EFO/v1><br><br>{<br>RECORD ?r {<br>?s RECATT ?ra {?p ?o} | {GRAPH <dataset_dictionary> {<br><EFO> evo:hasInstantiation <EFO/v1> .<br><EFO/v1> evo:hasRecordSet ?rs<br>} GRAPH ?rs{<br>?r a evo:Record ; evo:subject ?s ;<br>evo:hasRecordAttribute ?ra . |
| }<br>} | ?ra evo:predicate ?p ;<br>evo:object ?o }} |
| FROM CHANGES <EFO> BETWEEN VERSIONS <EFO/v1> <EFO/v2><br><br>{<br>CHANGE ?c {?p ?o}<br>} | {GRAPH <dataset_dictionary> {<br>?cs a evo:ChangeSet ;<br>evo:oldVersion <EFO/v1> ;<br>evo:newVersion <EFO/v2><br>} GRAPH ?cs{<br>?c a _:Change ; ?p ?o }} |

# 6 Evaluation

In this section we present the evaluation of our approach over a real world evolving biological use case of the EFO ontology as well as use case concerning evolving multidimensional data of the statistical domain published on the web in LOD format following the Data Cube Vocabulary[9] approach. As a first step, in Table 7, we provide a qualitative evaluation of supported storage policies, querying scopes, supported change representation, and metadata granularity of a framework implementing the DIACHRON model and Query language, compared with related works discussed in Section 2. Specifically, we compare our approach with traditional version control, as well as SemVersion [32], Auer and Herre [1], Im et al [8], Hauptmann et al [7] and Memento LD [30,31], and we find that these approaches cover parts of the functionality offered by a framework that implements DIACHRON. Furthermore, we conducted a performance evaluation and a usability evaluation. The performance evaluation aims at showing that there is no significant overhead imposed in query processing that introduces above-linear performance for queries of increasing difficulty. The usability evaluation aims at measuring with objective metrics the syntax overhead that the proposed DIACHRON Query Language introduces.

In the first case, we consider 15 consecutive versions of the ontology, that exhibit various types of changes, both simple and complex, as well as four multidimensional datasets each comprised of three consecutive versions. We load all datasets into the same archive instance, and in order to do so, the data are first converted to fit the RDF mapping of the DIACHRON model. For this, we implemented a conversion mechanism as part of the Data Modeller component presented in the previous section. The modeller reifies data to records and record attributes. Data are mapped to the DIACHRON data model in the following manner. First, classes and their definitions (domains, ranges) are modelled as schema objects. The triples are grouped by their subjects. For each subject URI, its corresponding predicate-object pairs are modelled as record attributes and grouped in records. The subject records are in turn connected with the record attributes created for each triple associated with a subject URI. For a more in-depth discussion of the mapping process the reader is referred to [17].

[9] http://www.w3.org/TR/vocab-data-cube/

**Table 5.** **Qualitative comparison of each framework's support for (a) storage policies, (b) querying scopes, (c) change representation, and (d) provenance and metadata granularity. (CB = change-based storage, FM = full materialization)**

| | Storage | Querying | Changes | Provenance Granularity |
|---|---|---|---|---|
| Version control | CB (sequential) | N/A | Low level | None |
| SemVersion | FM | Graph Patterns | Low level | None |
| Auer et al | CB (sequential) | Changes | High level | Changes |
| Im et al | CB (aggregated) | Graph Patterns | Low level | Datasets |
| Hauptmann et al | CB (sequential) | Graph Patterns | Low level | Datasets |
| Memento LD | FM | Resources | N/A | Resources |
| **DIACHRON** | **Hybrid** | **Datasets, Versions, Graph Patterns, Resources, Changes, Longitudinal** | **High level** | **Datasets, Versions, Resources, Changes, Triples** |

### 6.1 Experimental Evaluation

The goal of the experimental evaluation was to assess the performance of our implementation w.r.t three main aspects: the time overhead related to the initial loading of the archive, the time overhead related to the retrieval of the datasets in their original form (de-reification and serialization) and the time overhead of executing queries of different difficulty. Specifically, we want to assess (i) the runtime performance of the pre-processing step for DIACHRON QL, and (ii) whether there is extra processing overhead that makes query processing non-linear with respect to query difficulty. Our approach was implemented in Java 1.7, and all experiments were performed on a server with Intel i7 3820 3.6GHz, running Debian with kernel version 3.2.0 and allocated memory of 8GB.

First, bulk operations on whole datasets have been tested, namely loading and retrieving full dataset versions. Loading and retrieval times can be seen in **Figure 9** (a) and (b). A series of 10 tests were run for each version of the datasets and the averages have been used in computing execution time, using least squared sums. Loading a dataset in the archive implies splitting it into the corresponding structures, i.e. dataset, record set, schema set and change set, and storing it in different named graphs. The splits were done directly in the store using the SPARQL update language and basic pattern matching, thus no need to put a whole dataset in memory arose, which would be costly in terms of loading in and building the respective Java objects in Jena[10]. The increasing sizes of the input datasets are the effect of their evolution, as new triples are being added. In the same **Figure 9** (b), retrieval times can be seen for the same datasets. Retrieval of a dataset is the process of de-reifying it to recreate the dataset version at its original form and structure. As can be seen, both loading times and retrieval fit into a linear regression w.r.t to the datasets' sizes as measured in record attributes and imply that no additional time overhead is imposed that would destroy linearity as new versions of a dataset are stored in the archive.

**Figure 9** (c)-(h) show running times of 14 queries we devised for this experiment. An analysis of the queries' characteristics can be seen in **Table 6**. In **Figure 9** (c) and (d) we perform a series of queries on different dataset versions. Specifically, two sets of 5 queries have been devised to run on a fixed dataset. Each query is run on one particular version, and the total running time of all 5 queries in each set (c) and (d) is calculated after retrieving the results and storing them in memory, which implies a simple iteration on all results. The query sets are made up from SELECT queries that combine structural entities (records, record attributes etc.) with actual data entries (subject URIs etc.) in different levels of complexity. In **Figure 9** (a) no aggregate functions, OPTIONALs or other complex querying capabilities have been used, while in **Figure 9** (b) the queries consist of selecting, aggregating and filtering graph patterns. As in the case of loading and retrieval, the archive behaves in a linear way as the size of a dataset increases.

Finally, four queries, Q11-Q14, with variable datasets that search in the entire archive have been devised and run on an incrementally larger archive, that is, the queries have been tested on deployments of the archive where versions of datasets are being incrementally added to their corresponding diachronic datasets. The queries use dataset versions as variables. The results can be seen in **Figure 9** (e)-(h) where linearity is still being preserved when new datasets are stored.

Running times for the pre-processing step can be seen in **Figure 10**. Specifically, we have measured the total running time required to create and populate a DIACHRON query object, prior to execution, as opposed to a SPARQL query object, for queries Q1-14 on an archive instance that contains the maximum number of tested versions. The pre-processing overhead for DIACHRON QL is proportional to the intermediate steps, but does not impose a large difference when compared with plain SPARQL queries in the majority of cases. The SPARQL queries appear to impose a constant overhead, while the time needed to pre-process DIACHRON queries increases along with the query expressivity and complexity of mapped scopes and DIACHRON elements. Even so, the pre-processing overhead is negligible (in most cases <100ms). For queries Q13 and Q14 the pre-processing step is very costly, because of the sequenced nature of pre-processing steps required to combine materialized and non-materialized datasets in queries with variable diachronic datasets and dataset instantiations.

[10] https://jena.apache.org/

**Table 6. Characteristics of the experiment queries**

| | Q1 | Q2 | Q3 | Q4 | Q5 | Q6 | Q7 | Q8 | Q9 | Q10 | Q11 | Q12 | Q13 | Q14 |
|---|---|---|---|---|---|---|---|---|---|---|---|---|---|---|
| DISTINCT | √ | √ | √ | √ | √ | √ | √ | √ | √ | √ | √ | √ | √ | √ |
| Unbound predicates | | | | | | | | √ | √ | √ | √ | √ | √ | √ |
| Filters | | | | | | | | | | | √ | √ | √ | √ |
| Aggregate Functions | | | | | | | √ | √ | √ | √ | √ | √ | √ | √ |
| ORDER BY | | | | | | | √ | √ | √ | √ | √ | √ | √ | √ |
| OPTIONAL | | | | | | | | | | | | √ | | |
| SELECT | √ | √ | √ | √ | √ | √ | √ | √ | √ | √ | | √ | √ | √ |
| CONSTRUCT | | | | | | | | | | | √ | | | |
| Reified data | √ | √ | √ | √ | √ | | | | | | | | √ | √ |
| De-reified pattern | | | | | | √ | √ | √ | √ | √ | √ | √ | √ | √ |
| Diachronic metadata | | | | | √ | √ | | | √ | √ | √ | √ | √ | √ |
| Unbound named graphs | | | | | | | | | | | √ | √ | √ | √ |
| Non-materialized datasets | | | | | | | | | | | | | √ | √ |

### 6.2 Usability Evaluation

In order to evaluate usability, we make use of three objective metrics in order to compare the *compactness*, *expressiveness* and *usability* of DIACHRON QL with respect to SPARQL. Specifically, we compare (i) the number of language-specific keywords used in each of the 14 queries, (ii) the total number of triple/record patterns, and (iii) the number of intermediate variables that were neither part of the original query, nor requested by the user. The results can be seen in Table 7.

As the number of SPARQL TPs increases, the number of DIACHRON QL record patterns remains at low levels, thus abstracting the complexity of writing large queries. This is especially evident in queries Q13 and Q14, where we have used hybrid storage policies, thus forcing the query engine to decide on parse-time which dataset versions are materialized and which have to be materialized as nested graph patterns. For instance, query 13 that features a bound diachronic dataset with an unbound version (using `AT_VERSION ?v`) can be expressed with just two DIACHRON patterns, whereas the SPARQL query uses 33 triple patterns to cater for the versions that follow a mixed storage policy. Note, however, that independently of the underlying storage policies, even if the user was inclined to express a query in a language like SPARQL and rely on an existing query engine for execution, the set of intermediate steps executed by our system would be omitted in the process, thus limiting the expressivity of the possible queries, as was discussed in section 5.

The number of keywords used in each of the two languages for the 14 queries is smaller for small queries (queries Q1-Q5), but SPARQL tends to overcome DIACHRON in total number of language-related keywords as the query gets larger and more complicated. This is also dependent on the various scopes and filters used by a query. Finally, SPARQL eventually depends on a number of dynamically generated intermediate variables that are used in the translated query, which is not needed by DIACHRON. These variables bind dictionary elements, scopes, versions, record sets and so on to variables that are further used in GRAPH clauses and FILTERs in the SPARQL translation.

**Table 7. Comparison of (i) number of keywords, (ii) number of triple (or record) patterns, and (iii) number of generated variables not existing in the original query.**

| | Q1 | Q2 | Q3 | Q4 | Q5 | Q6 | Q7 | Q8 | Q9 | Q10 | Q11 | Q12 | Q13 | Q14 |
|---|---|---|---|---|---|---|---|---|---|---|---|---|---|---|
| # keywords (SPARQL) | **4** | **4** | **4** | **4** | **4** | 7 | 9 | 8 | 8 | 8 | 10 | 12 | 27 | 46 |
| # keywords (DIACHRON) | 6 | 6 | 5 | 5 | 5 | **5** | **5** | **6** | **6** | **5** | **7** | **7** | **6** | **7** |
| | | | | | | | | | | | | | | |
| # TPs (SPARQL) | 5 | 5 | 5 | 5 | 10 | 6 | 7 | 9 | 9 | 9 | 19 | 21 | 33 | 44 |
| # TPs (DIACHRON) | **1** | **1** | **1** | **1** | **1** | **2** | **2** | **2** | **2** | **2** | **4** | **4** | **2** | **2** |
| | | | | | | | | | | | | | | |
| # non-TP vars (SPARQL) | 2 | 2 | 1 | 2 | 3 | 2 | 2 | 3 | 2 | 2 | 4 | 5 | 6 | 9 |
| # non-TP vars (DIACHRON) | **0** | **0** | **0** | **0** | **0** | **0** | **0** | **0** | **0** | **0** | **0** | **0** | **0** | **0** |

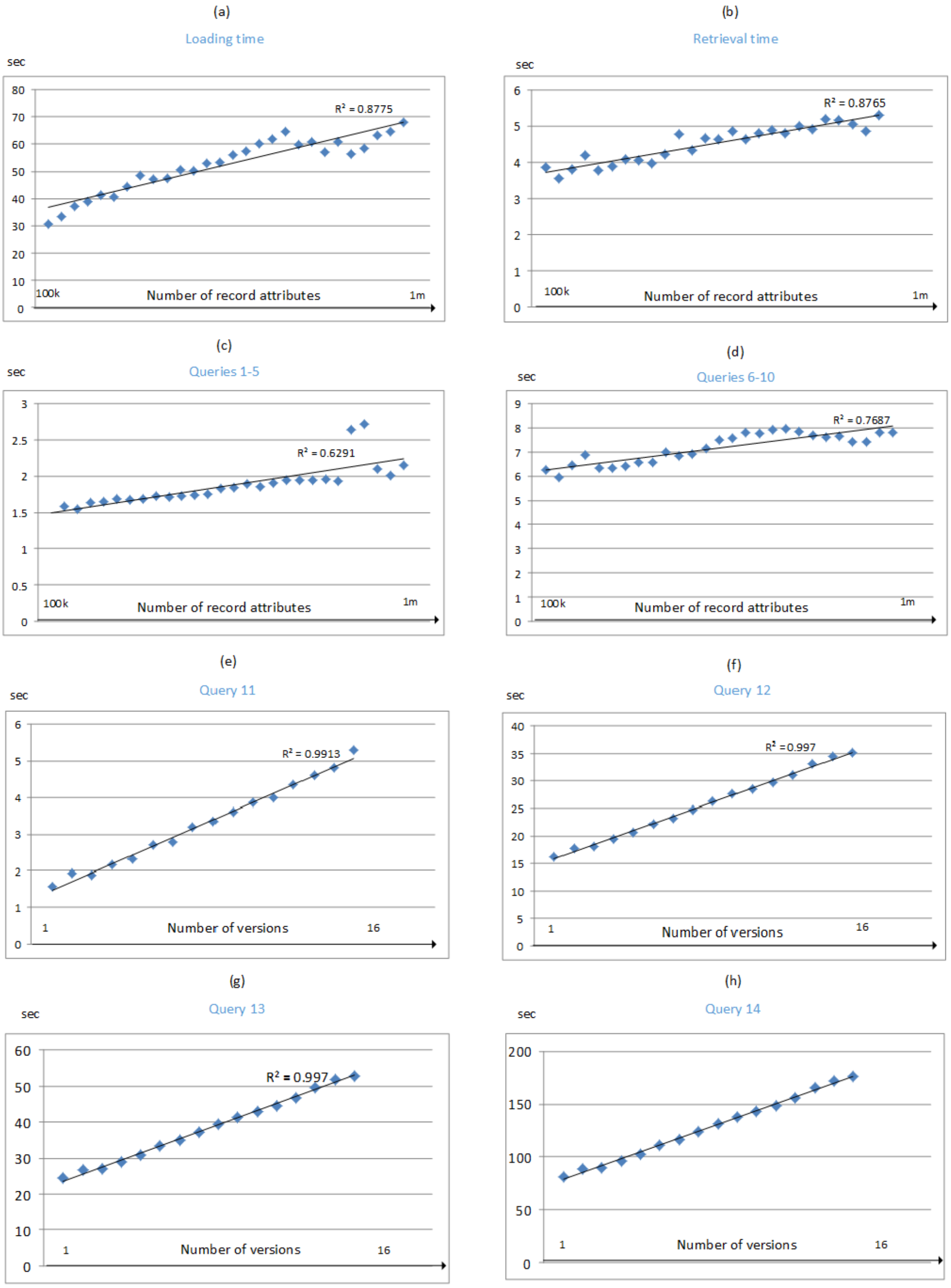

Figure 9: Loading times (a), retrieval times (b), select queries without filters and aggregates (c), select queries with filters and aggregates (d), select queries with variable datasets (e)-(h).

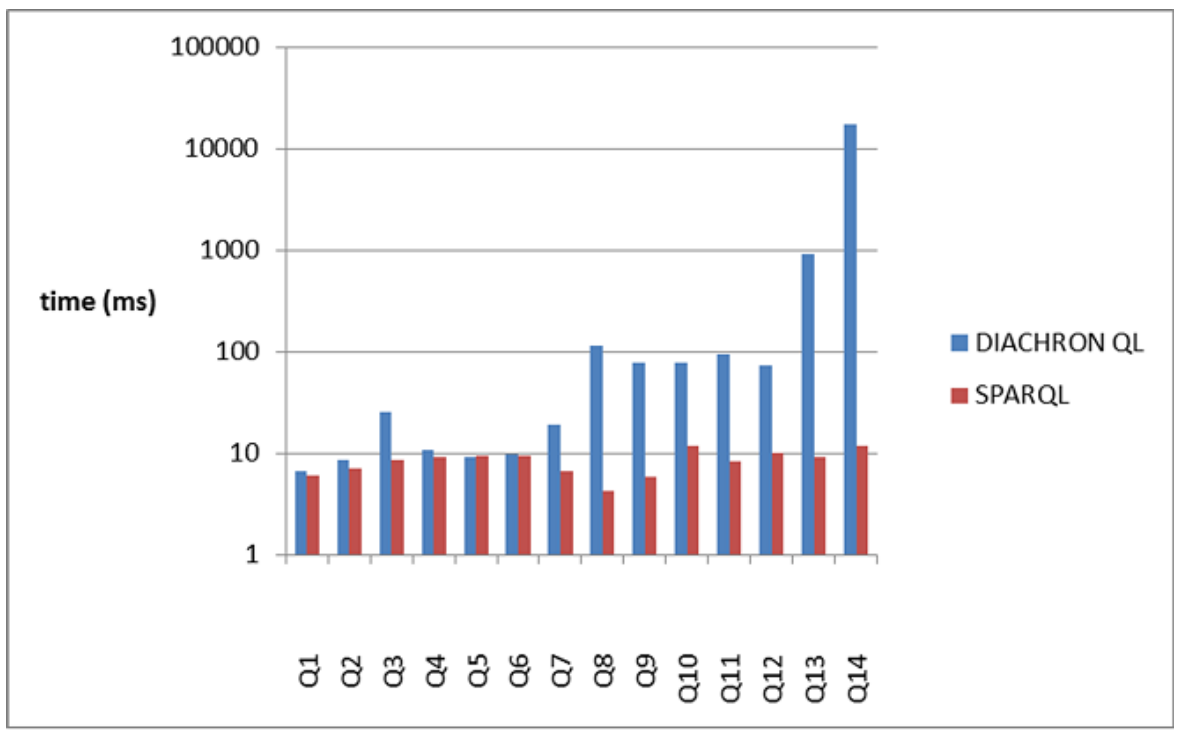

Figure 10. Logarithmic plot of pre-processing time (in milliseconds) for queries Q1-Q14.

# 7 Conclusions

In this paper, we have discussed the challenges and requirements for the preservation and evolution management of datasets published on the Data Web and we have presented an archiving approach that utilizes a novel conceptual model and query language for storing and querying evolving heterogeneous datasets and their metadata. The DIACHRON data model and QL have been applied to real world datasets from the life-sciences and open government statistical data domains. An archive that employs these ideas has been implemented and its performance has been tested using real versions of datasets from the aforementioned domains over a series of loading, retrieval and querying operations.

The growing availability of open linked datasets has brought forth significant new problems related to the distributed nature and decentralized evolution of LOD and has posed the need for novel efficient solutions for dealing with these problems. In this respect, we have highlighted some possible directions and presented our work that tackles evolution and captures several dimensions regarding the management of evolving information resources on the Data Web.